\documentclass{aa}   

\usepackage{natbib}
\usepackage{graphicx}
\usepackage{txfonts}
\usepackage{hyperref}

\nolinenumbers
\usepackage[normalem]{ulem}

\begin{document}

   \title{Revisiting the Milky Way stellar long bar and the 3 kpc arm\thanks{
   {\bf In memoriam of Terence J. Mahoney.} The authors dedicate this paper to the memory of our 
   late colleague, Terence J. Mahoney (1949--2025), who worked with some of us (MLC and FGL) 
   over the last 30 years on analyses of the long bar within the Two-Micron Galactic Survey (TMGS) team (Tenerife, Spain).}}
   \subtitle{}
   \author{M. L\'opez-Corredoira$^{1,2}$, W. Wu$^{3,4,1,2}$, 
   H.-F. Wang$^{5,6,7}$, F. Garz\'on$^{2,1}$}
   
   \offprints{martin@lopez-corredoira.com}

\institute{$^1$ Instituto de Astrof\'\i sica de Canarias, E-38205 La Laguna, Tenerife, Spain\\
$^2$ Departamento de Astrof\'\i sica, Universidad de La Laguna, E-38206 La Laguna, Tenerife, Spain \\
$^3$ Institute of Astronomy and Physics, Inner Mongolia University, Hohhot 010021, China \\
$^4$ National Astronomical Observatories, Chinese Academy of Sciences, Beijing 100101, China \\
$^5$ Local Universe and Time-Domain Astronomy Laboratory, China West Normal University, Nanchong 637002, China\\
$^6$ Luoxiahong Institute of Astronomy, China West Normal University, Nanchong 637002, China \\
$^7$ Department of Astronomy, China West Normal University, Nanchong 637002, China
}

   \date{}

  
  \abstract
{One of the most difficult and unexplored regions of the Milky Way is the highly extincted in-plane central region within
the Galactic coordinates $10^\circ \lesssim |\ell |\lesssim 30^\circ $, $|b|\lesssim 3^\circ $, where we have
the long-bar and 3 kpc arm with intermediate-age stellar population, whose morphological properties are still unclear.}
{We aim to advance our knowledge of the morphology of these two components.}
{We examined star counts of bright M giants in WISE-4.6$\mu $m and its distribution of distances
derived from spectroscopic parallaxes with APOGEE-DR17. We also examined the distribution of distances of young 
OGLE-O-rich Mira variable stars, and reviewed the literature on red clump distance determination within that area.}
{We corroborate the asymmetry between positive and negative longitudes in in-plane regions, thus confirming the necessity to include
a long bar. We obtain an average angle between the major axis of the long bar and the line Sun--Galactic centre of $\alpha =27.4^\circ \pm 1.5^\circ $, aligned with the triaxial bulge and a semi-major-axis length $\approx 4$ kpc.
The tips of the long bar are in contact with the elliptical 3 kpc arm, with the major axis again aligned with the bulge and the long bar's major axes, whose tangential lines of sight correspond to $\ell =-22^\circ $ and $\ell=+27^\circ $. In the range of 50 degrees in the sky between these two longitudes, 
the stellar near 3 kpc arm is clearly detected at heliocentric distances around 5 kpc, and the stellar far 3 kpc arm is tentatively detected at heliocentric distances of 9-12 kpc.}
{}

\titlerunning{Long-bar and the 3 kpc arm}
\authorrunning{L\'opez-Corredoira et al.}

\maketitle
\nolinenumbers
%

\section{Introduction}
\label{.intro}

One of the most difficult and unexplored regions of the Milky Way is the in-plane central region, namely within
the Galactic coordinates $|\ell |\lesssim 30^\circ $, $|b|\lesssim 3^\circ $. On the one hand, the extinction of the stars 
within central volume is high.
On the other hand, the complexity of the Galactic structure is high, with several components to disentangle, which
makes it difficult to analyse given that we observe them edge-on.
For the stellar populations, apart from the stellar halo and disc and spiral arms, 
we have a triaxial bulge (dominant within $|\ell |\lesssim 10^\circ $) 
and other components necessary to explain the observed 
asymmetry of the star counts within $10^\circ \lesssim |\ell |\lesssim 30^\circ $. 
Two major components have been identified for the stellar in-plane central regions so far: 
the long bar and the 3 kpc arm (or inner ring).

\citet{Dam86} pointed out the existence of a large-scale
long bar in the CO map whose tip is at the beginning of the Scutum arm. A similar conclusion examining stars in the near-IR band was
obtained by \citet{Ham94}, which is the origin of the hypothesis of a stellar long bar. Its geometry was investigated
either through the identification of the possible bar tips \citep{Lop01,Lop07,Gon12,Amo13} or by measuring the distances
of some of its stars, mainly with red clump stars \citep[e.g.][see further references and
the discussion in Sect. \ref{.rc}]{Ham00,Ben05,Cab08,Weg15} or other standard candles \citep[e.g.][]{van03}.
This long bar has stars younger (age$\lesssim $ 6 Gyr) 
than those of the bulge at a larger distance from the plane \citep{Ng96,Col02,Nie26}
and metallicities that are disc-like or higher \citep{Gon08,Weg19}, which reveals either a peculiar wings-like extension of the bulge \citep[a thicker structure; e.g.][]{Mar11} or a structure different from the bulge with a different origin and/or evolution \citep{Lop11}. In any case, the existence of a bulge and a long bar in a spiral galaxy is not surprising.
There are also a few external galaxies that exhibit a bar and boxy bulge both visible photometrically \citep{Sel93,Bet94,Com14} or with kinematical signatures in edge-on galaxies \citep{Kui96}.

One of the features of the long bar subject to dissensus is its orientation:
it is claimed by some works to be the same as or very slightly larger than the triaxial bulge one ($\alpha $ between 20 and 30 degrees) or much larger by other authors.
Using red clumps' standard candles, \citet{Weg15} gave $\alpha =28-33^\circ,$ 
where $\alpha $ is the angle between the major axis and the Sun--Galactic centre line, whereas much larger values
were obtained with the red clumps by other works: $\alpha >40^\circ$ \citep[e.g.][]{Ham00,Cab08}.
Using \textit{Gaia} Data Release (DR) 2 data, \citet{And19} gave an angle of the bulge+long bar of $\alpha \approx 45$ deg, 
with 30 million stars with photometry and parallaxes alone at $|Z|<3$ kpc. It is dominated by the bulge in $|\ell |<10^\circ $
regions, but it includes in-plane regions for larger $\ell $, showing the long bar with the angle of 45 deg.
\citep[Fig. 8]{And19}. However, apart from the huge errors of the parallax for sources at distances larger than 5 kpc, {\it Gaia} wavelength is significantly affected by extinction, so it is not the most appropriate survey to study the long bar.
\citet{Que21} considered that the angle of \citet{And19} 
was overestimated because of extinction saturated over $A_V=4$ and the uses of parallaxes alone. \citet{Que21} used 
stars from \textit{Gaia} DR2 and the Apache Point Observatory Galactic Evolution Experiment (APOGEE) DR16 with better distance determinations that have $\sim $10\% errors and 0.2 kpc$<|Z|<$1 kpc, combined with other surveys, and obtained an angle value of $\sim $20 deg, though this was more focused in the bulge areas than the long-bar region.
\citet{And22} reconsidered the calculations with improved photo-astrometric distances and stellar density maps using \textit{Gaia} Early DR3 data and multi-band photometry. Their StarHorse code includes a prior assumption of the orientation angle of the bulge and the long bar of 27$^\circ $, while their density map based on the red clumps suggests a bar angle that is a few degrees larger than the priors, especially near the tips of the long bar \citep[Fig. 8]{And22}. The analysis does not separate the bulge from the long bar, and it uses some priors, which makes it model dependent.

The question of the possible misalignment of the bulge and long bar remains unclear.
Also, from the 3D decomposition of some galaxies with a long bar (e.g. NGC 3992), a significant 
misalignment between the long bar and the bulge of $\Delta \alpha >20^\circ $ was claimed \citep{Com14}.
Two triaxial structures (plus other components) with different angles is not theoretically impossible;
they might be oscillating systems governed by a periodically time-dependent gravitational potential of variable length bars \citep{Abr86,Lou88,Gar14}, or their size ratio may be extreme, as in some double-barred galaxies \citep{She09}. 
However, in most present-day models of bar formation, the two structures are presented as aligned or with only a
slight twist near the tips due to an interaction with the adjacent spiral arm heads (see \citealt{Mar11} and references therein),
understanding the existence of a bulge and a long bar as different components of the same coherent bar structure as seen in simulations \citep{Ath05,Mar11,Li15}. Other dynamical aspects \citep[e.g.][]{Bla16,She20,Luc23,Hun25,Che25} 
have been associated with the features of the inner Galaxy, though most dynamical studies
focus on the bulge (usually called the `bar') and do not explicitly calculate the effect of this extension within the plane
known as the `long bar'.

In the outskirts of the long-bar sweeping area, we find another structure called the 3 kpc arm. Historically, it was called the `arm',
though its nature is not related to the spiral arms. 
Rather than a spiral arm, the stars of the 3 kpc arm constitute an 
elliptical ring that delineates a trail of quasi-elliptical (X1) orbits 
around the Galactic bulge and long bar \citep{Kum25}. 
The 3 kpc arm structure appears likely to correspond to the radius of co-rotation resonance of the long bar \citep{Gre11,Bla16}, 
with the long bar on its inner surface and the starting points of the spiral arms on its outer surface.
The near part of the 3 kpc arm was discovered long time ago in HI \citep{van57}, and CO evidence
came later for the far 3 kpc arm behind the bulge and long bar \citep{Dam08}. It includes young stars identified by maser emission
\citep{Gre11,Kum25} and a predominantly old stellar population, with an average age of $\sim 6$ Gyr \citep{Wyl22};
the vertical thickness of the ring decreases markedly towards younger ages \citep{Wyl22}. There are red clump giants with high metallicity \citep{Bla16}, and asymptotic giant branch stars \citep{Chu09}. \citet{Dam08} show that the 3 kpc arm contains a significant amount of molecular gas. The fact that this dense gas is not associated with strong tracers of recent star formation (such as HII regions; \citealt{Loc80}) is a key piece of evidence in support of suppressed star formation in a region with a high concentration of gas, thus giving few or no OB main-sequence stars. It is likely due to the gas being highly turbulent and shocked by the Galactic bulge and long bar. 
The long bar gradually transitions into a radially thick, vertically thin, elongated inner ring with an average solar [Fe/H] \citep{Wyl22}.

This stellar elliptical ring, called the 3 kpc arm, is so named because of its most prominent detection at $\ell =-22^\circ $, $b=0$
\citep{Gre11,Dav12}, which corresponds to a tangential line of sight; assuming a Sun--Galactic centre distance of $d\approx 8$ kpc gives $R=d\,|\sin(-22^\circ )|\approx 3$ kpc, though the radius is variable because it is not a perfect circular ring.  
\citet{Mik82} and \citet{Rue91} interpreted the $\ell=27^\circ $, $b=0$ flux excess as the other tangential line of sight of the 3 kpc arm 
in positive longitudes, though $\ell=27^\circ $ was also interpreted as the possible tip of the long bar connecting to
the Scutum spiral arm \citep{Ham94,Lop99,Lop01}. This interpretational dissensus is also related to the  angle, $\alpha, $ of the long bar, because a larger value of $\ell $ for the tip in the positive longitudes favours a larger
$\alpha $.

In this work, we aimed to shed further light on the morphology and nature of the long bar and 3 kpc arm.
Section \ref{.WISE} of this paper describes how we used near- to mid-IR data of the Wide-field Infrared Survey Explorer (WISE; \citealt{Wri10}) 
to show evidence of the asymmetry
of star counts between the positive and the negative in-plane within $10^\circ \lesssim|\ell |\lesssim 30^\circ $, which reinforces
previous analyses even with a low extinction filter and even when a correction of extinction is included.
In Sect. \ref{.APOGEE} we describe how we used spectroscopic parallaxes of APOGEE-2 (\citealt{Maj17}) data, with a distance accuracy of 10--20\%, to derive the angle of the long bar.
Section \ref{.OGLE} outlines our use of Mira variable stars from the Optical Gravitational Lensing Experiment (OGLE) survey \citep{Iwa22} 
with a relative distance precision of 4\% to trace the structure of the long bar; we provide a new determination of its angle 
and the 3 kpc arm morphology.
Section \ref{.rc} features a reanalysis of the literature concerning the red clump's relationship to the long bar and searches for the possible reasons of
dissensus.
Section \ref{.discuss} provides a summary and discussion of the results.

\section{In-plane long-bar with WISE photometry}
\label{.WISE}

\subsection{WISE data}
\label{.WISEdata}

The National Aeronautics and Space Administration (NASA)'s WISE (\citealt{Wri10}) mapped the sky at 3.4, 4.6, 12, and 22 $\mu $m (W1, W2, W3, and W4) with an angular resolution of 6.1", 6.4", 6.5", and 12.0" in the four bands. According to the description on its web site, 
the AllWISE data \citep{Cut14} we used here combine data from the WISE cryogenic and Near-Earth Object Wide-field Infrared Survey Explorer (NEOWISE) \citep{Mai11} post-cryogenic survey phases to form the most comprehensive view of the full mid-IR sky currently available. By combining the data from two complete sky-coverage epochs using an advanced data-processing system, AllWISE generated new products that have enhanced photometric sensitivity and accuracy and improved astrometric precision compared to the 2012 WISE All-Sky Data Release. Exploiting the 6--12 month baseline between the WISE sky coverage epochs enables AllWISE to measure source motions for the first time and to compute improved flux-variability statistics. AllWISE contains accurate positions, apparent motion measurements, four-band fluxes, flux-variability statistics, and cross-correlations with Two-Micron All-Sky Survey (2MASS) sources (point-like and extended) for over 747 million objects detected on the co-added Atlas Images. 

Photometry was performed using point-source profile-fitting and multi-aperture photometry. WISE 5$\sigma $ photometric sensitivity is estimated to be at 16.6, 15.6, 11.3, and 8.0 Vega mag at 3.4, 4.6, 12, and 22 $\mu $m in unconfused regions on the ecliptic plane. Sensitivity is better at higher ecliptic latitudes where coverage is deeper and the zodiacal background is lower, and poorer when limited by confusion in high-source-density or complex-background regions. Saturation affects photometry for sources brighter than approximately 8.1, 6.7, 3.8, and -0.4 mag at 3.4, 4.6, 12, and 22 $\mu $m, respectively.

Here, we removed the extended sources (included in 2MASS-XSC extended-source catalogue), the sources with variability (flag ivafl1$\ge 6$) and those with a signal-to-noise ratio lower than 3. We limited our analysis to the Galactic coordinates 
$|\ell |<40^\circ $, $|b|<5^\circ $: a total of 19\,136\,150 point-like sources, 
612\,417 of them with $m_{4.6\mu m}<8.0$.

These sources have associated J, H, and K$_s$ magnitudes from cross-correlation with 2MASS.
In 2MASS/J-filter, the limiting magnitude of completeness is 15.8 mag, reducing it to 14-15 mag in the crowded fields; this was expected in the plane of the galaxies. This means that we were limited to complete counts when $(J-W2)\lesssim 6-7$, which may affect the loss
of very reddened sources, but these are very few;
for instance, of the 125\,995 sources with $m_{4.6\mu m}<8.0$, $|b|\le 0.5^\circ$ (the in-plane region for the relevant analysis of the long bar) only 172 have no counterpart with all filters of 2MASS. 
The cross-correlation might have errors given the high crowdedness of these fields and the low angular
resolution of WISE, but for very bright sources the number of in-plane sources is relatively low (around 2\,000 deg$^{-2}$), and the probability of misidentification is low.

\begin{figure}[htb]
\vspace{1cm}
{
\par\centering \resizebox*{8.5cm}{7cm}{\includegraphics{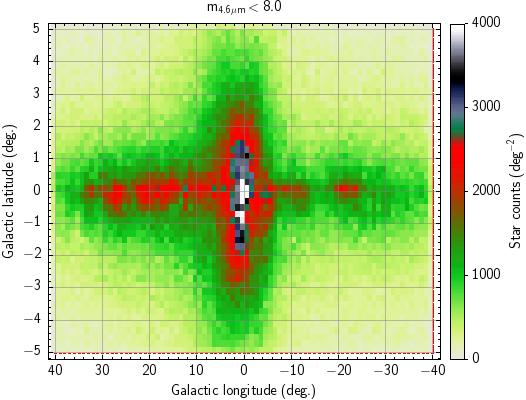}}
\par\centering
\par\centering \resizebox*{8.5cm}{7cm}{\includegraphics{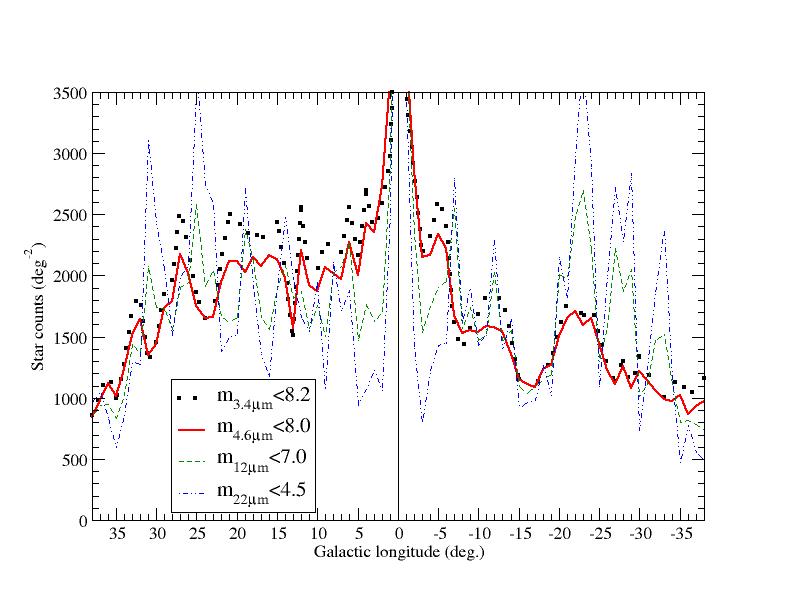}}
\par\centering
}
\caption{WISE star counts. Top: As a function of $\ell$, $b$ within $|\ell |<40^\circ $, $|b|<5^\circ $, $m_{4.6\mu m}<8.0$, $\Delta \ell =1^\circ $, $\Delta b=0.1^\circ $.
Bottom: As a function of $\ell $ within $|\ell |<40^\circ $, $|b|<0.5^\circ $, for $m_{3.4\mu m}<8.2$, $m_{4.6\mu m}<8.0$,
$m_{12\mu m}<7.0$, $m_{22\mu m}<4.5$, $\Delta \ell =1^\circ $.}
\label{Fig:WISEcounts}
\end{figure}

\begin{figure}[htb]
\vspace{1cm}
{
\par\centering \resizebox*{8.5cm}{7cm}{\includegraphics{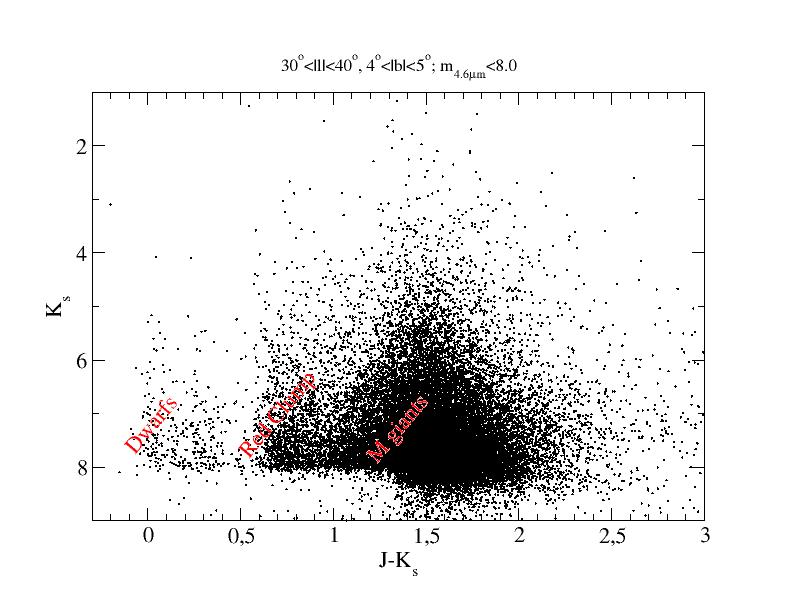}}
\par\centering
}
\caption{Colour-magnitude diagram in low extincted areas ($30^\circ <|\ell |<40^\circ $, $4^\circ <|b|<5^\circ $;
$\langle E(J-K)\rangle =0.4$; \citealt{Sch98}) for sources with $m_{4.6\mu m}<8.0$.}
\label{Fig:CMDJK}
\end{figure}

\begin{figure}
\vspace{0.2cm}
{
\par\centering \resizebox*{8cm}{6.3cm}{\includegraphics{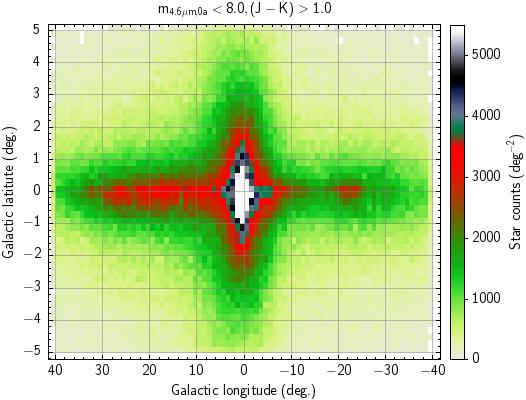}}
\par\centering
\par\centering \resizebox*{8cm}{6.3cm}{\includegraphics{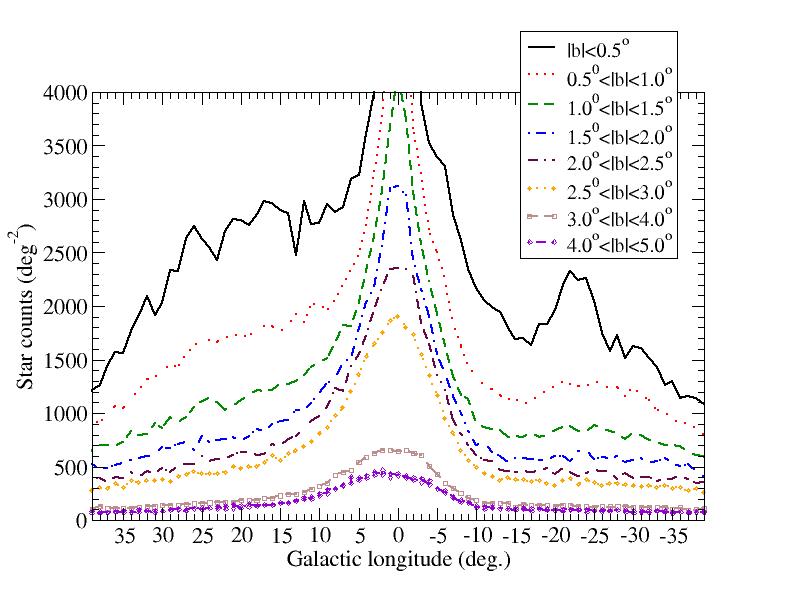}}
\par\centering
\par\centering \resizebox*{8cm}{6.3cm}{\includegraphics{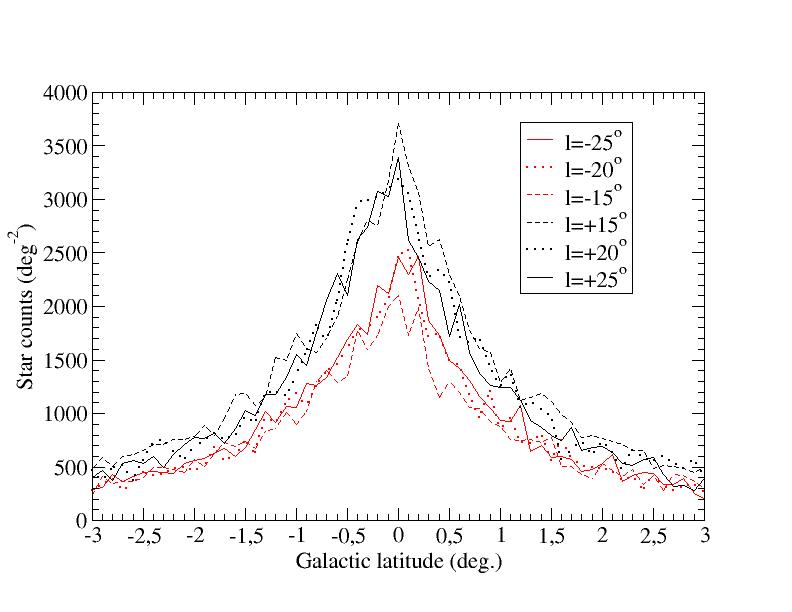}}
\par\centering
}
\caption{WISE star counts as in Fig. \ref{Fig:WISEcounts} (top panel) but with stars with $J-K<1.0$ removed (dwarfs and red clumps from Galactic disc) and with the correction of extinction through Eq. (\ref{extcorr}). 
We observe a prominent asymmetry between positive and
negative Galactic longitudes and a very slight asymmetry between positive and negative Galactic latitudes.
}
\label{Fig:WISEcountsce}
\end{figure}

\subsection{Asymmetry in star counts}

In Fig. \ref{Fig:WISEcounts} we show the star counts within $m_{4.6\mu m}<8.0$. We clearly see an asymmetry within $|b|<0.5^\circ $,
with further counts for positive than negative longitudes within $5 \lesssim |\ell |\lesssim 28^\circ $. This asymmetry is gradually attenuated for larger values of $|b|$.

As said in the introduction, this asymmetry has been interpreted in terms of the existence of a new component of the Galaxy, namely long bar, which is different from the (thick) bar or bulge; it is also observable here and with its own asymmetry for off-plane regions.
Previous analyses of this asymmetry \citep{Ham94,Lop01,Lop07,Weg15} have shown it in the K band, Galactic Legacy Infrared Midplane Survey Extraordinaire (GLIMPSE) 3.6$\mu $m, GLIMPSE 4.5$\mu $m, and the Midcourse Space Experiment
(MSX) 8$\mu $m; here, we show the same trend with WISE 4.6$\mu m$.. Other bands of WISE show a similar trend (Fig. \ref{Fig:WISEcounts}, bottom panel).

The 3 kpc arm (inner ring) without a long bar does not explain the in-plane features \citep{Ham94,Lop01}. A ring would be expected to produce a peak in the counts at the positions tangential to the line of sight: a U-shape rather than the monotonously increasing counts with $\ell $ within $10^\circ \lesssim \ell \lesssim 27^\circ $. If the ring were elliptical, the longitudes of the peaks would no longer be symmetric; however, the shapes of the peaks would remain basically unaltered. It is possible that a contrived ring (particularly a patchy ring), coupled with a highly improbable distribution of extinction, could reproduce the form of the in-plane counts. However, as we show here, this distribution of extinction does not exist, and hence by itself the ring cannot explain the in-plane counts.

Extinction in 4.6$\mu $m is very low: $A_{4.6\mu m}=0.026A_V=0.33A_K=0.082E(B-V)=0.158E(J-K)$ \citep{Wan19}. 
The application of an extinction correction cannot be carried out with a 3D extinction map, because we have no information on the
distances of the sources here. The cumulative extinction cannot be applied for all stellar populations, because
it affects the sources in the centre of the Galaxy behind the dust disc, but not nearby sources in the disc.
Nonetheless, we can carry out an approximate correction for the extinction by means of
\begin{equation}
\label{extcorr}
A_{4.6\mu m(W2)}(\ell, b)=\langle 0.158\,[(J-K)-1.1]\rangle _{\ell ,b,(J-K)>1.0}
\end{equation}\[
m_{4.6\mu m(W2),0a}\equiv m_{4.6\mu m}-0.158\,[(J-K)-1.1];\] 
\[ \ \ \ \ \ \ \ \ \ \ \ \ \ \ {\rm only\ for\ (J-K)>1.0.} 
\]
The value of intrinsic colour, $(J-K)_0=1.1,$ is derived as the average colour (1.5) in regions within our selected area of low extinction; this is shown in Fig. \ref{Fig:CMDJK} for $(J-K)>1.0$ stars, with a correction of $\langle E(J-K)\rangle=0.4$ 
in this region \citep{Sch98}. This intrinsic color $(J-K)_0 =1.1$ corresponds to M giants \citep{Cov07}. 
Given the range of selected magnitudes ($m_{4.6\mu m}<8.0$), we are dominated by M giants at heliocentric distances of $d=$5-12 kpc, 
with some contamination
of dwarfs and red clumps at $d<1$ kpc from the disc. With the constraint $(J-K)>1.0,$ we removed most of the main-sequence stars and red clump giants of the disc (see Fig. \ref{Fig:CMDJK}). Different combinations of filters other than J and K are not as precise when it comes to the correct identification and separation of disc stars.
We know this method is not perfect; as stated in Sect. \ref{.WISEdata}, we lose some stars in very
extincted regions which reach $J>14-15$, but these are very few (negligible). We applied average extinction laws, which may have
some fluctuations. Nonetheless, for our purpose, it is good enough.

\begin{figure}[htb]
\vspace{1cm}
{
\par\centering \resizebox*{8.5cm}{7cm}{\includegraphics{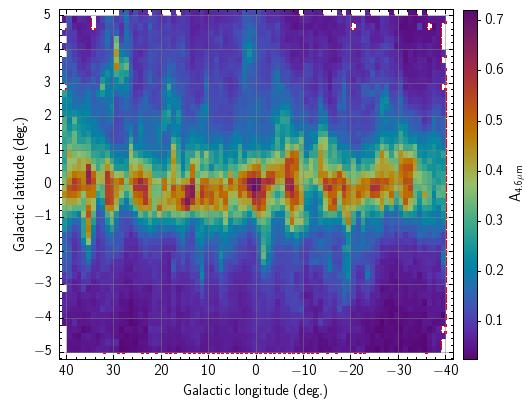}}
\par\centering
}
\caption{Extinction $A_{4.6\mu m}$ as a function of $\ell$, $b$ within $|\ell |<40^\circ $, $|b|<5^\circ $; $\Delta \ell =1^\circ $, $\Delta b=0.1^\circ $, derived from Eq. (\ref{extcorr}).}
\label{Fig:ext4p6}
\end{figure}

Figure \ref{Fig:WISEcountsce} gives the map of extinction-corrected star counts: $m_{4.6\mu m,0a}>8.0$, $(J-K)>1.0$. We observe the same asymmetry of star counts in in-plane regions than in Fig. \ref{Fig:WISEcounts}, though with much smaller fluctuations for both the bulge and the long bar. Figure \ref{Fig:ext4p6} shows the extinction in 4.6$\mu $m, which is low, as expected: around 0.5 magnitudes on average within $|b|<0.5^\circ $ and with a small dependence on the Galactic longitude. There is not an asymmetry in extinction between positive
and negative Galactic longitudes, apart from fluctuations and a small tilt in the dependence with $b$ at $|\ell |\lesssim -15^\circ $. Everything is in agreement with previous observations by \citet{Lop01}. 

The conclusions are the same as those of \citet{Ham94} and \citet{Lop01,Lop07}: extinction cannot explain the asymmetry we observe in star counts. Neither the disc nor spiral arms can produce this asymmetry. Hence, there should be a structure of stars (here, we observe mainly M giants, but there are other populations) that are well constrained within the plane and have strong non-axisymmetry; these are in the long bar. The observed in-plane excess of stars located at $\ell \approx -22^\circ $ corresponds to a
young massive star cluster \citep{Gre11,Dav12}
and is thought to be a tangential cross of the 3 kpc arm \citep{Lop01}. 

The excess of star counts at positive longitudes with respect to negative longitudes for $15^\circ <|\ell |<30^\circ$  
indicates that the closest part of the long bar is in the first quadrant. This asymmetry of star counts is observed even at $|b|\gtrsim 3^\circ $ (Fig. \ref{Fig:WISEcountsce}, middle panel), though the very fast decline of star counts with increasing latitude
indicates that the structure is much more concentrated in the plane than the bulge (=thick bar; dominant at $\ell \lesssim 10^\circ $).

There is also a slight asymmetry in the vertical direction, with further star counts at negative latitudes than positive latitudes
(Fig. \ref{Fig:WISEcountsce}, bottom panel). The star counts are vertically symmetrical with respect to $b=-0.2^\circ $ instead of $b=0$.
This small shift might be due some residuals of extinction correction, but it most likely has to do with the position of the Sun over the Galactic plane. A shift of $z_\odot \sim +20$ pc over the plane (as observed; e.g. \citet{Kar17} estimated that the Sun lies 
at $z_\odot =17\pm 5$ pc above the Galactic mid-plane, and the median of 55 previous estimates 
published is $z_\odot =17\pm 2$ pc) means that a structure at $d\approx 6$ kpc is vertically 
shifted $\sin ^{-1} \left(\frac{-z_\odot}{d}\right)\approx -0.2^\circ $ with respect to $b=0$.

These WISE stars have, on average, a corrected-for-extinction H-magnitude equal to $H_0\approx 7.5$, which at distances of $<5$ kpc corresponds to absolute magnitudes of $M_H<-6.0$, which again is in agreement with them being M giants \citep{Cov07}. 
We see the distribution of distances to these M-giant
stars in Sect. \ref{.APOGEE}.

\subsection{Angle of the long bar}

In the first quadrant, the tip of the long bar is typically situated at $\ell =27^\circ -30^\circ$ \citep{Lop99,Lop01,Lop07,Zha14}.
Here, we also see a fast drop of counts at exactly $\ell =27^\circ $ in Fig. \ref{Fig:WISEcounts} (bottom panel), in agreement
with that idea. However, $\ell =27^\circ $ might instead be a peak due to the tangential cut of the 3 kpc arm \citep{Mik82,Rue91},
and the tip of the long bar would be placed in a lower Galactic longitude.
Examining the continuity of the excess of stars in the in-plane region with $\ell>0$, a more
conservative position for the tip of the long bar in the positive longitudes would be $\ell _1=27^\circ \pm 3^\circ$, given that
we also observe a significant drop of stars at $\ell =24^\circ $ even after extinction correction (Fig. \ref{Fig:WISEcountsce}, top panel) and that the region of excess stars might extend up to $\ell \approx 30^\circ $.

Conversely, in the negative longitudes, the terminus of the long bar is located between $\ell =-12^\circ $ and $\ell =-14^\circ $ \citep{Lop01,Gon12,Amo13}. Here,in  examining the extinction-corrected star counts of Fig. \ref{Fig:WISEcountsce} (top), we would
place the limits as $\ell _2=-12^\circ \pm 2^\circ $.

Given these data, with a projection of the long bar, we can derive the angle of the bar ($\alpha $)
 with respect to Sun--Galactic centre's line of sight. In fact, it is a simple trigonometrical problem if we take the bar to be rectilinear and of equal length ($L_0$) from each end to the centre \citep{Lop99}:
\begin{equation}
\frac{L_0}{\sin (\ell _1)}=\frac{R_\odot}{\sin (\alpha +\ell_1)}
,\end{equation}\[
\frac{L_0}{\sin (-\ell _2)}=\frac{R_\odot}{\sin(\alpha +\ell _2)}
,\]
where $R_\odot$ is the Sun--Galactic centre distance. Hence,
\begin{equation}
\label{barangletips}
\alpha =\tan ^{-1}\left(\frac{-2}{\cot (\ell _1)+\cot (\ell _2)}\right)
\end{equation}
(independent of $R_\odot$). With the above values of $\ell _1$ and $\ell _2$, we obtain\begin{equation}
\label{WISEalfa}
\alpha =36^{+13}_{-9}\ {\rm deg.}
\end{equation}\[
L_0=(0.51\pm 0.08)R_\odot
.\]

\section{In-plane long bar with APOGEE-2 spectroscopy}
\label{.APOGEE}

\subsection{APOGEE-2 data}

We used the Sloan Digital Sky Survey Data fourth-phase (SDSS-IV) DR17 (\citealt{Abd22}) as a source for the following: the H-band (1.51-1.70 
$\mu $m) high-resolution ($R\sim 22,500$) spectra 
of the APOGEE-2 survey, with distances derived by \citet{Sto24} for 733,901 independent sources with near-IR spectra using neural networks with uncertainties of 10-20\%.
The data include northern- and southern-hemisphere observations from two different telescopes and spectrographs:
APOGEE spectrograph on the 2.5m Sloan Foundation Telescope at Apache Point Observatory (APO) in New Mexico, USA and
an identical spectrograph on the 2.5m Ir\'en\'ee du Pont Telescope at Las Campanas Observatory (LCO) in Chile.
The Galactic disc is targeted with a more or less systematic grid of pointings within $|b|<15$ deg. For $0<\ell <30$ deg., 
there is denser coverage of the bulge and inner Galaxy. It provides a statistically robust sample for studying the Galaxy, 
but it does not contain every star in its magnitude range.

The typical H-band magnitude range was $\approx 7$ to $\approx 13.5$.
APOGEE-2's primary sample is magnitude-limited ($H<12.2$, or $H<12.8$ in some bulge fields; using 2MASS photometry), 
but its completeness is heavily modified by a density-based selection algorithm \citep{Zas17}.
This means that while the pool of potential targets is all stars brighter than a certain H-band magnitude, not all of these stars are observed. The selection is designed to handle the immense stellar density, particularly toward the Galactic bulge and disc.
APOGEE-2 does not randomly select all stars brighter than H=12.2. Instead, it uses a statistical sampling rate that varies with stellar density.
In low-density fields (e.g. high Galactic latitude and halo), the sampling rate is 100\%. 
In high-density fields (e.g. the Galactic plane and bulge), the probability of a star being selected is inversely proportional to the local stellar density. As a result, the completeness as a function of magnitude in a crowded sightline is nearly flat for bright magnitudes (e.g. from H=7.0 to H=12.2), but at a value of much less than 100\%. The drop in completeness is not primarily due to magnitude, but to density on the sky. 
Furthermore, with the `Faint Time' and Ancillary Science Samples, APOGEE-2 also included secondary targets,
in the $12.2<H<13.5$ range (sometimes even fainter for special programs like star clusters or dwarf galaxies).
APOGEE-2's main target selection also used colour cuts: $(J-K)_0>0.5$ in disk and bulge areas, which means that the sources are mainly K or M spectral types, though there are also some sources with bluer colours.

\begin{figure}
\vspace{1cm}
{
\par\centering \resizebox*{8.5cm}{7cm}{\includegraphics{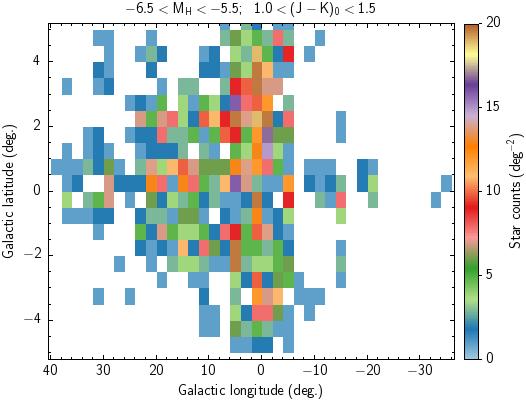}}
\par\centering
\par\centering \resizebox*{8.5cm}{7cm}{\includegraphics{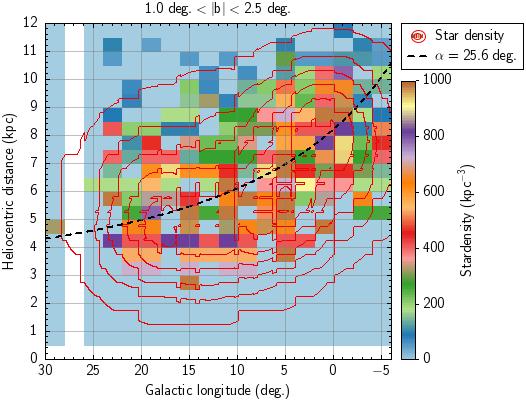}}
\par\centering
}
\caption{Top: APOGEE-2/DR17 M5-6III ($-6.5<M_H<-5.5$, $1.0<(J-K)_0<1.5$) star counts, 
in the range of Galactic coordinates  $|\ell |<40^\circ $, $|b|<5^\circ $. These counts are highly in incomplete (and white
areas are unobserved regions) due to the observing strategy of APOGEE-2, 
but for each line of sight the completeness is constant, with a magnitude
within $7.0<H<12.2$ when $A_H<2.7$. 
Bottom: Distribution of distances of these stars for $-6^\circ <\ell <30^\circ $, $1.0^\circ <|b|<2.5^\circ $;
dashed line shows the best fit to a long bar with an angle of $\alpha =25.6^\circ$ for $\ell>10^\circ $.}
\label{Fig:counts}
\end{figure}

We are only interested in the central in-plane areas of the Milky Way, so we added the following constraints: $|\ell |\le 40^\circ $,
$|b|\le 5^\circ $. 
Here, we select a subsample with the criteria
$-6.5<M_H<-5.5$, $1.0<(J-K)_0<1.5$
within a distance of $5<d<10$ kpc and $A_H<2.7$; they correspond to $7.0<H<12.2$, which is within
the range of observed magnitudes, where there is flat completeness as a function of magnitude.
This range of absolute magnitudes and colours corresponds to M5-6III (M giants) stellar types \citep{Cov07}.
In total, we have 1\,391 stars with these characteristics; see Fig. \ref{Fig:counts} (top panel).
This selection of stars should contain a negligible amount of disc stars for distances larger than 12 kpc (M5-6 giants
are too faint there) or lower than 4 kpc (they are too bright).

The deficit of stars in the innermost plane ($|b|\lesssim 1^\circ $) is due to their higher extinction, with
many stars having $A_H>2.7$.
We should avoid regions with high extinction: not only should the average $A_H$ be much lower than 2.7 
(equivalent to $A_{4.6\mu m}<0.51$ at Fig. \ref{Fig:ext4p6}), but also,
due to the patchiness of the extinction, the ratio of stars with $A_H>2.7$ should be low ($<5$\% of the stars at a distance
between 5 and 10 kpc). This is followed for lines of sight of
$|b|>1^\circ $, so we apply this constraint in the following analyses.

\subsection{Distribution of distances}

With these APOGEE data, we do not see the asymmetry positive--negative longitudes as clearly as with WISE data because there are very few data with negative longitude in the in-plane regions. Furthermore, the scarcity of APOGEE sources is strongly concentrated at $\ell >5$, which makes the top panel of Fig. \ref{Fig:counts} rather incomplete. Nonetheless, APOGEE provides distances, whereas WISE does not, which allowed us to see a significant variation of heliocentric distances with Galactic longitude; the latter was used to determine the angle of the long bar with more precision than when only using the positions of the tips of the bar in WISE. The bottom panel of Fig. \ref{Fig:counts} contains a sufficient number of stars per bin to accurately determine the location of the density peak along each line of sight, thereby allowing the bar's position angle to be reliably traced.

Figure \ref{Fig:counts} (bottom panel) shows the distribution of distances for $-6^\circ <\ell <30^\circ $, $1^\circ<|b|<2.5^\circ $
(a total of 1\,060 stars). The star density is calculated as $\rho =\frac{N(r)dr}{\omega r^2\,dr}$, where $\omega $ is the solid angle of each bin and $N(r)dr$ is the star counts between $r-dr/2$ and $r+dr/2$.
We chose this range of Galactic longitudes to trace the long bar, avoiding its far side because our
sample does not reach the very distant stars in the negative Galactic longitudes. We avoided the areas of highest extinction with
$|b|<1^\circ $. On the other hand, $|b|$ was restricted to the region where we find a significant presence of the long bar in the WISE survey.

Figure \ref{Fig:counts} (bottom panel) shows a continuous structure with a slightly decreasing distance for an increasing Galactic longitude.
This indicates the presence of a barred structure up to $\ell=+26^\circ $. 
The dispersion of the distance values is of the order of the distance error; thus, 
the long bar should not be a thick structure. The average distance of the central stars ($|\ell|<2^\circ $) is 8.3 kpc.
Note that the disk contribution in the central regions is not significantly detected; for instance, at $|\ell|<5^\circ $ we
observe few stars with $d<5$ kpc, and for $15^\circ <\ell <30^\circ $ we find very few stars with $d>10$ kpc. 
There is not an axisymmetric structure in the central areas for these selected stars of Fig. \ref{Fig:counts} (bottom panel).
Therefore, we neglected the contribution of the disk.

\subsection{Angle of the long bar}
\label{.APOGEEangle}

Given that the APOGEE survey is highly incomplete, we were not able to build a 3D map of the density in the Galactic-bar region. 
A fit of the angle of the long bar in an incomplete 3D map would be very complex and subject to many uncertainties with regard to the
selection function.
Nonetheless, keeping in mind that for each line of sight the distribution of distances $d$ 
between $\approx 5$ and $\approx 10$ kpc has the same level of completeness, a direct fit of the long-bar structure of 
$d(\ell )$ is possible. For each line of sight with a given $\ell, $ we crossed the long bar at a certain distance, $d_L$, and
the number of stars with measured $d$ over and below $d_L$ should be similar.
Assuming the long bar is a narrow structure and the maximum density of stars in the line of sight traces its major axis (see Appendix A of  
\citealt{Lop07}), the distance of this structure in the in-plane regions is
\begin{equation}
\label{d_l}
d_L(\ell )=R_\odot \frac{1}{\cos \ell +(\cot \alpha \sin \ell )} 
,\end{equation}
where $\alpha $ is the angle of the long bar with respect to the lines Sun--Galactic centre. We assumed $R_\odot =8.2\pm 0.2$ kpc
(from different measurements in the literature). 
We also assumed all of the stars in the line of sight are at the same distance.
This approximation is not valid for a (thick) bulge, so the analysis of a triaxial bulge angle cannot be carried out
with the same simplistic assumptions.

With our distribution of distances, $d(\ell ),$ for the 167 stars with
$10^\circ <\ell <30^\circ $ (we avoided the region dominated by the thick bulge) and $1.0^\circ <|b|<2.5^\circ $ (we avoided the high-extinction areas and limited the maximum latitude where the asymmetry in WISE star counts is significant), 
a minimum $\chi ^2$ analysis gives a value of $\alpha =25.6^\circ \pm 2.3^\circ $.
The dashed line in the bottom of Fig. \ref{Fig:counts} shows this best fit.
The errors in the fitting include global statistical errors and are due to the assumed error of $R_\odot$.
For $10^\circ <\ell <24^\circ $ (avoiding the thick bulge and possible confluence with the 3 kpc arm), $1.0^\circ <|b|<2.5^\circ $:
$\alpha =25.5^\circ \pm 2.3^\circ $ (165 stars).

\subsection{With StarHorse distances}

\begin{figure}
\vspace{1cm}
{
\par\centering \resizebox*{8.5cm}{7cm}{\includegraphics{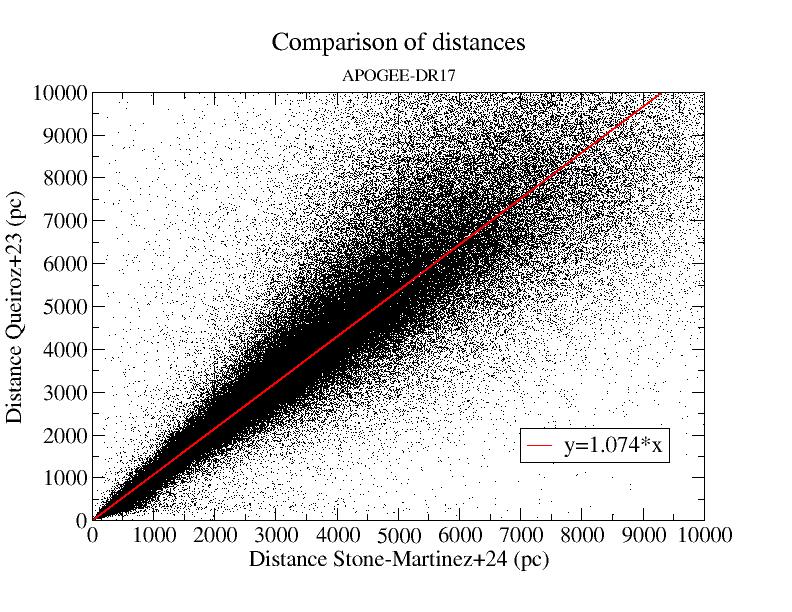}}
\par\centering
}
\caption{Comparison of distances estimated for APOGEE-DR17 with the algorithms of \citet{Que23} and \citet{Sto24}, respectively.}
\label{Fig:QueirozvxStone}
\end{figure}

We used distances from \citet{Sto24} in APOGEE-DR17 stars. There is another estimation of distances of APOGEE-DR17 sources
using the StarHorse algorithm \citep[only for 562\,424 stars;][]{Que23}, which gives an average of 7.4\% larger distances for stars within a Stone-M.-distance of $<10$ kpc (see Fig. \ref{Fig:QueirozvxStone}). 
From Eq. (\ref{d_l}), $\cot \alpha_{StarHorse}=\frac{\cot \alpha _{Stone-M.}-0.074\cot \langle \ell \rangle }{1.074}$.
For $\alpha _{Stone-M.}=25.5^\circ $, $\langle \ell \rangle=17^\circ $, we obtain 
$\alpha _{StarHorse}=30.1^\circ $. 

\section{In-plane long bar with OGLE Mira variable stars}
\label{.OGLE}

\subsection{OGLE data of Mira variable stars}

The \citet{Iwa22} catalogue contains 65\,981 Mira stars in the Milky Way based on the third and the fourth phases of the OGLE project. There are two subtypes: C-rich and O-rich.
From the total, 40\,356 stars are from bulge fields and 25\,625 Miras are in disc fields. 
The authors provided light curves in the I and V bands from the Johnson--Cousins photometric system collected from 1996, December to 2020, March. The collection covers $\sim 3000$ deg$^2$ of the sky. The catalogue containing equatorial coordinates, pulsation periods ($P$), I-band brightness amplitudes, mean magnitudes in the V and I bands, light curves, and finding charts is publicly available and can be accessed through the OGLE Internet Archive.\footnote{https://ogle.astrouw.edu.pl/} 
The relative precision of the distance determination is around 4\% \citep{Cat16}.
Cross-correlation with mid-IR data including WISE filters was added by \citet{Iwa23}. \citet{Iwa23} limited their exploration space to a 3D cuboid: -4.2 kpc$\le X\le$3.8 kpc, 
-4 kpc$\le Y \le$4 kpc, and -2 kpc $\le Z\le $ 2 kpc, with (X,Y,Z) Cartesian Galactic coordinates centred in the Galactic centre
($R_\odot =8.2$ kpc) and thus covering the central parts of the Milky Way. This limit left us with 39\,619 Miras, which are the ones used here.

\begin{figure}
\vspace{1cm}
{
\par\centering \resizebox*{8.5cm}{7cm}{\includegraphics{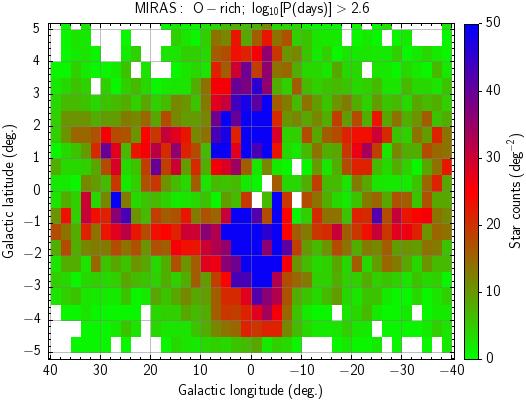}}
\par\centering
\par\centering \resizebox*{8.5cm}{7cm}{\includegraphics{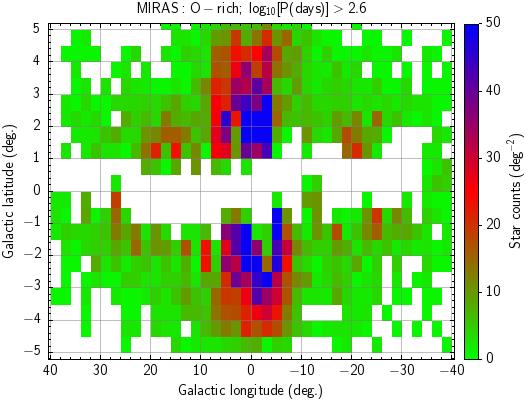}}
\par\centering
}
\caption{Top: O-rich Mira variable star counts, with a period, $P,$ within $\log_{10}[P({\rm days})]>2.6$, 
in the range of the Galactic coordinates  $|\ell |<40^\circ $, $|b|<5^\circ $. 
Bottom: Same as  the top panel but with the stars within the mask corresponding to regions of high extinction and low
completeness removed (see \citealt{Iwa23}).}
\label{Fig:MIRAScounts}
\end{figure}

In this section we restrict our analysis to the Miras within $|\ell |\le 40^\circ $, $|b|\le 5^\circ $. 
We also removed C-rich Miras, as these stars usually change their mean brightness over time due to the significant mass-loss phenomenon \citep{Iwa23}. We are interested in the brightest
and youngest population, which offer a better contrast between the long bar and the bulge, so we only selected stars with $\log _{10}[P({\rm days})]>2.6$; they correspond to $M_{4.6\mu m}<-8.5$ (apparent W2 magnitude between 5.0 and 6.5 at distances
between 5 and 10 kpc with negligible extinction; \citealt{Iwa21}) and ages $<5$ Gyr \citep{Cat16}.
With all of these criteria, there remains 10\,581 O-rich Mira stars. Figure \ref{Fig:MIRAScounts} (top panel) shows the star-count
distribution as a function of Galactic coordinates.

The completeness of these sources is much lower than 100\% in regions with high extinction (in I-filter), as indicated by
\citet{Iwa23} with a mask. Outside the mask, in areas where completeness is almost 100\%, there are 5\,371 O-rich Mira stars.
Figure \ref{Fig:MIRAScounts} (bottom panel) shows the star counts once the stars in the masked regions are removed.
The comparison of masked and unmasked regions shows that we lose stars due to incompleteness even at $|b|$ in the entire region,
though predominantly at $|b|<1^\circ $. The loss of stars of bins at higher latitudes is due to the existence of a few (not all) small areas with high extinction, even when the average extinction is not high due to the patchiness of the extinction.

\subsection{Distribution of distances}

\begin{figure}
\vspace{1cm}
{
\par\centering \resizebox*{8.5cm}{7cm}{\includegraphics{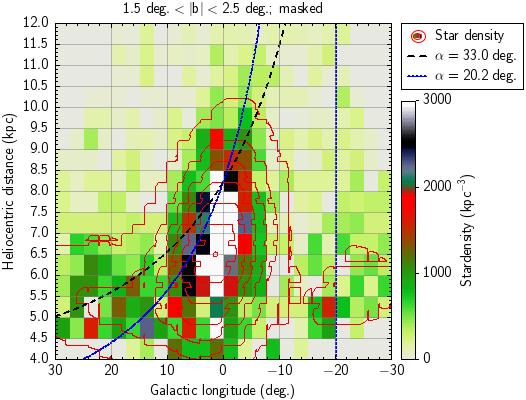}}
\par\centering
\par\centering \resizebox*{8.5cm}{7cm}{\includegraphics{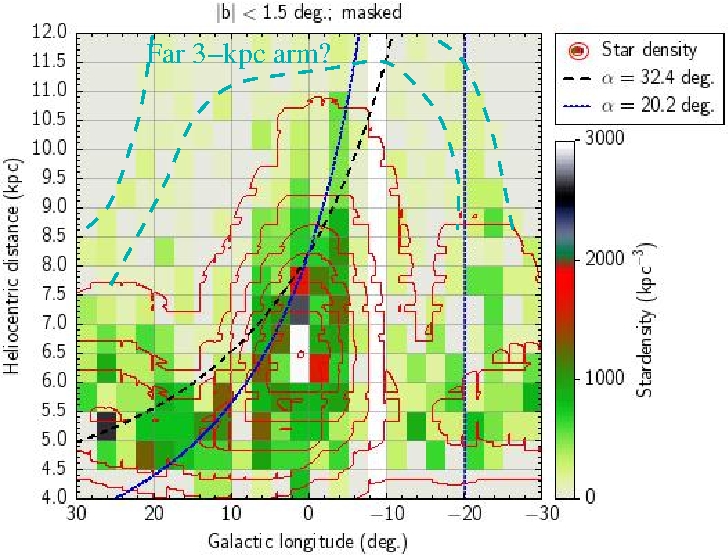}}
\par\centering
}
\caption{Distribution of distances of O-rich Mira variable stars 
with a period, $P,$ within $\log_{10}[P({\rm days})]>2.6$
for $-20^\circ <\ell <30^\circ $, $1.5^\circ <|b|<2.5^\circ $ (top panel; moderate extinction, high completeness) or $|b|<1.5^\circ $ (bottom panel; high extinction, low completeness).
The dashed line shows the best fit to a long bar within $10^\circ <\ell<30^\circ $  for bins with a mask (removing the high-extinction areas), $d<R_\odot $, whereas the dotted line shows the corresponding $d(\ell )$ for an angle of $\alpha =20.2^\circ $, as fitted for the major axis of the bulge \citep{Iwa23}.
The dashed cyan line marks a tentative imprint of the far 3 kpc arm.}
\label{Fig:MIRASdist}
\end{figure}

In Fig. \ref{Fig:MIRASdist} (top panel) we show the distribution of distances for stars with a mask (removing the high-extinction areas)
for $1.5^\circ <|b|<2.5^\circ $, where the extinction is moderate on average.
In Fig. \ref{Fig:MIRASdist} (bottom panel), we show the distribution of distances for stars with mask 
for $|b|<1.5^\circ $, where the extinction is high, on average, and the completeness is low, except in a small portion of areas 
corresponding to holes of extinction.

The bulge area $|\ell |<10^\circ $ is well represented by the global angle of the major of the bulge.
\citet{Iwa23} calculated an angle of the major axis of the bulge $\alpha =20.1^\circ, $ and, although there should be some
difference between the maximum star-count line and the major axis due to the thickness of the bulge \citep[Appendix A]{Lop07}, we see that this angle may also roughly represent the bulge region in near-plane regions. 
However, there is a prolongation
between $\ell =10^\circ $ and $\ell =30^\circ $ that belongs to another structure with a different angle.
The distance obtained at $\ell =25-30^\circ $ is around 5.5 kpc, which agrees with the value obtained by \citet{Zha14} of 
$5.49^{+0.39}_{-0.34}$ kpc. This feature also resembles the distribution given by \citet[Fig. 6]{Weg15} for red clump
stars in the K band.

\begin{figure}
\vspace{1cm}
{
\par\centering \resizebox*{8.5cm}{7cm}{\includegraphics{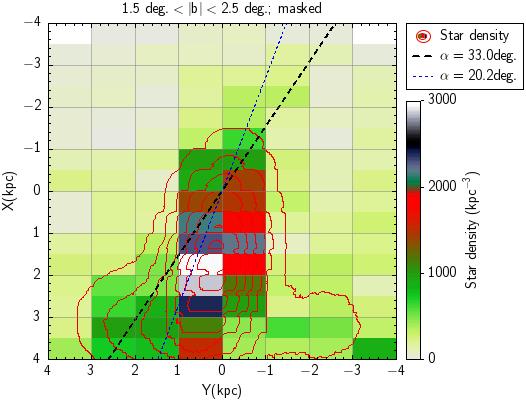}}
\par\centering
\par\centering \resizebox*{8.5cm}{7cm}{\includegraphics{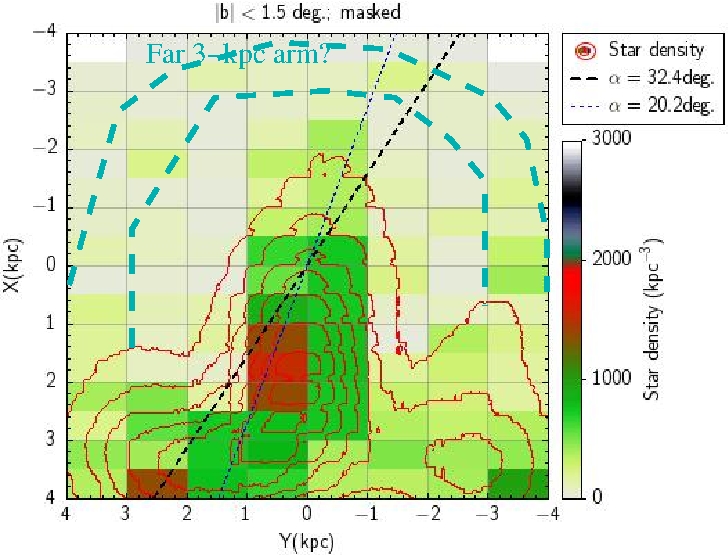}}
\par\centering
}
\caption{Same stars as Fig. \ref{Fig:MIRASdist} with Cartesian coordinates.}
\label{Fig:MIRAScart}
\end{figure}

Given the incompleteness in some bins, especially at $|b|<1.5^\circ $, the deficit of stars at some Galactic longitudes cannot
be interpreted in terms of density drops. Nonetheless, since we used the stars outside the mask of low-incompleteness areas, the distribution of distance as a function of Galactic longitude should not be affected. 

The shape of the isodensity contours might be interpreted as a long-bar detection at $10^\circ \lesssim \ell \lesssim 24^\circ $,
though with some deficit of stars due to incompleteness.
The strong excess of stars at $\ell \approx 24-30^\circ$ corresponds to a well-known in-plane region where the star formation rate is high \citep{Lop99,Neg10,Zha14}, with a constant heliocentric distance of $d\approx 5.5$ kpc; this is significantly deviated from 
the extrapolation of the bar at shorter wavelengths. This may indicate a structure that is not a straight bar, or
easier to understand in terms of the fragment of the 3 kpc arm, which is tentatively the same one tangentially observed at $\ell \approx -22^\circ, $ but at positive longitudes.

\subsection{Angle of the long bar}
\label{.Mirasalfa}

With the same type of calculation provided in Sect. \ref{.APOGEEangle} for bins in the masked map, within $10^\circ <\ell<30^\circ $, $d<R_\odot $ and for $1.5^\circ <|b|<2.5^\circ $  (211 stars) we obtain an average value of $\alpha =33.0^\circ \pm 2.2^\circ $.
For $|b|<1.5^\circ $  (174 stars), we obtain a value of $\alpha =32.4^\circ \pm 2.2^\circ $; the error bars include global statistical errors and a result of the assumed error of $R_\odot=8.2\pm 0.2$ kpc.

In Fig. \ref{Fig:MIRASdist} we see that a single component of the long bar with fixed angle does not reproduce all of the data. We think this is due to the fact there is an extra component (3 kpc arm or stellar ring) for larger Galactic longitudes of 20--30 deg.). The combination of the long bar and 3 kpc arm has a large average 
$\alpha $ value along the lines of sight near the tip of the bar, which is likely to be at $\ell=24^\circ $, and beyond, where the ring dominates the counts.
Within $10^\circ <\ell<24^\circ $, $d<R_\odot $ for $|b|<2.5^\circ $  (262 stars; avoiding the region of possible confluence with the stellar ring), we obtain an average value of $\alpha =28.8^\circ \pm 2.0^\circ $.
Figure \ref{Fig:MIRAScart} shows the same stars as Fig. \ref{Fig:MIRASdist} but in Cartesian coordinates: $X=R_\odot-d\,\cos(\ell)$; $Y=d\,\sin(\ell)$ (we approximate $\cos(b)=1$).

\subsection{Far 3 kpc arm}

{Furthermore, a tentative detection of the far 3 kpc arm (the side of the ring beyond the Galactic bulge and long bar) is
observed in Fig. \ref{Fig:MIRASdist} (also observed in Fig. \ref{Fig:MIRASdist}, but less clearly due to
its lower space resolution at far distances).
The low signal precludes a firm confirmation of this detection, and further research
is warranted. Nonetheless, given the detection of the near 3 kpc arm with $d=4.0-6.0$ kpc, it is expected that the ring
is complete, as in the radio observations in CO \citep{Dam08}.
The observed density of the putative far 3 kpc arm is lower than the near 3 kpc arm, but this might be due to the larger
$\langle Z\rangle $ observed for lines of sight with constant $b$, given that $Z=d\sin(b)$.
In any case, our data only show the geometric distribution of some stars, with an overdensity in the regions associated with the 3 kpc arm previously observed, especially with gas, interpreted as the detection of the stellar component of the 3 kpc arm. We did not carry out an analysis of populations, ages, metallicities, and so on, which could have corroborated that this population does not belong to the disc.

\subsection{Geometrical parameters of an elliptical 3 kpc arm}

We can obtain the three parameters of an ellipse (major axis, minor axis, and angle) via the characterisation of three points.
With the observed tangential points at $\ell _1=+27^\circ \pm 1^\circ $, $\ell _2=-22^\circ \pm 1^\circ $; the heliocentric distance
$d=4.5\pm 0.5$ kpc at $\ell=0$; and $R_\odot =8.2\pm 0.2$ kpc, we determined the parameters
of the ellipse defining the 3 kpc arm. Following the equations in Appendix \ref{.ellipse}, we obtained a best fit for
\begin{equation}
\alpha =21^\circ \pm 3^\circ
,\end{equation}\[
a=(4.0\pm 0.4)\ {\rm kpc}
,\]\[
b=(3.1\pm 0.2)\ {\rm kpc}
.\]
The proximity of the value of the inclination, $\alpha, $ with that of the long bar (and the bulge) led us to consider
that the same $\alpha $ for the three structures is the most likely scenario.

\section{Review of the in-plane long bar with the red clump as a distance indicator}
\label{.rc}

As mentioned in Sect. \ref{.intro}, the existence of a long bar and the measurement of its angle 
has been mainly supported by the use of red clump giants.
Good analyses have already been carried out, and we cannot add anything new with respect to them.
Therefore, we simply discuss some information from previous papers to summarize the achievements.

Among the measurements of the long bar angle using red clumps as standard candles for in-plane stars, there are three groups of results:
\begin{itemize}
\item $\alpha $ between 20 and 25 degrees  \citep{Bab05}.
Although this work examined the in-plane regions, but only within $\ell <10^\circ $, which is dominated by bulge (thick bar) stars.
Therefore, we can discard this result.

\item $\alpha $ between 25 and 35 degrees   \citep{Zas12,Weg15}.

\item $\alpha $ between 35 and 45 degrees  \citep{Ham00,Ben05,Cab07,Cab08,Zas12a,Dur16}.

\end{itemize}

The discrepancy between the second group and the third group is due to two factors. The first is the different calibrations of the absolute magnitude of the red clump. For instance, in the mid-IR,
while \citet{Ben05} obtained a best
fit for $\alpha =44^\circ $ using $M_{4.5\mu m}=-2.15$, \citet{Zas12} used the same GLIMPSE data to obtain a best fit
for $\alpha =35^\circ $ using $M_{4.5\mu m}=-1.58$. The angle of the long bar is quite sensitive to the choice of the absolute magnitude of the red clump; fainter absolute magnitudes for the red clump give a shorter distance and, consequently, a lower angle.
A more recent calibration of the red clump's absolute magnitude gave \citep{Ple20} $M_{4.5\mu m}=-1.61\pm 0.29$ for low $\alpha $
enhancement or $M_{4.5\mu m}=-1.74\pm 0.22$ for high $\alpha $ enhancement, which favours the values given by \citet{Zas12}. 

Although the absolute magnitude of the red clump is generally thought to be constant, it actually varies with many factors, especially for metallicity and stellar age \citep{Gir01,Sal02,Bil13,Che17,Ono25}. Using seismically identified red clumps in the Kepler field, \citet{Che17} found that distance calibrations can be off by up to $\sim 0.2$ mag in the IR (over the range from 2 to 12 Gyr) if ages of these red clumps are unknown. Based on the red clumps selected from the Large Sky Area Multi-Object Fiber Spectroscopic Telescope (LAMOST) survey, 
\citet{Yu25} find that metal-poor red clumps ([Fe/H]$<-0.2$) could be fainter by 0.1--0.4 mag than the metal-rich counterparts in the Ks band. Our analysis is constrained to the Galactic plane of $|b|\lesssim 2^\circ $ in the inner Galaxy, which means that most of the selected red clumps should be metal-richer than -0.4 dex. There is no significant variation ($<0.1$ mag) in the absolute magnitude of red clumps with [Fe/H]$>-0.4$ \citep{Yu25}. Therefore, the metallicity dependence of the absolute magnitude can be neglected in this study.

The second factor is the  determination of distance from the maximum star counts versus magnitude [$N(m)$] for each line of sight in the third group, instead of maximum density versus distance $[\rho (r)]$ in the second group.
The last option is better, assuming there is negligible dependence of the density with $z$ (distance from the plane).
 Taking into account that $\rho (r)\propto \frac{N(m)}{r^3}$, the difference between two maxima is significant when there is some
significant dispersion of apparent magnitudes for the red clump, which is a volume effect \citep{Mar11}.
This difference was neglected by \citet{Ham00}, \citet{Ben05}, \citet{Cab07,Cab08}, \citet{Zas12a}, and \citet{Dur16}.

\citet[Appendix B]{Lop07} provided a method to take this difference between the two maxima into account,\footnote{There is an error
in \citet[Eq. B5]{Lop07}; it should be $\Delta r_m=\frac{3\rho (r_m)}{r_m\rho ''(r_m^*)}$ instead of 
$\Delta r_m=\frac{3\rho (r_m)}{r_m\rho ''(r_m)}$. In any case, this does not change the
estimation of $\Delta r_m$ when $\sigma <<1$.} which was used by \citet{Cab07} to claim that the difference of distance
between two maxima is only 25--50 pc (i.e. negligible). However, this calculation is not correct, 
and the number is much larger (of the order of 1 kpc). This shows that the long bar derived from $N(m)$ distribution is placed $\sim 1$ kpc away from the estimation using $\rho (r)$; consequently, it makes $\alpha $ larger.

This can be illustrated as follows: assuming $N(m)$ has a Gaussian distribution around its maximum with an rms $\sigma $,
\begin{equation}
\label{Gaus}
\rho (r)\propto \frac{1}{r^3}\exp {\left[-\frac{m(r)-m(r_m)}{2\sigma ^2}\right]}
,\end{equation}
with a low $\sigma $ approximation of $m(r)\approx m(r_m)+\frac{5}{\ln (10)}\left(\frac{r}{r_m}-1\right)$; hence,
the distance ($r_m^*$) maximum of the density distribution obtained when we set $\rho '(r_m^*)=0$ is
\begin{equation}
\label{correction}
r_m^*\approx r_m\,(1-0.64\sigma ^2)
.\end{equation}
For a typical $\sigma=0.5$ mag obtained by \cite{Cab07}, $\Delta r_m=r_m^*-r_m=-0.16\,r_m$, i.e. around 1 kpc for $r_m\approx 6$ kpc. This is consistent with the general expression of \citep[Appendix B]{Lop07} $\Delta r_m=\frac{3\rho (r_m)}{r_m\rho ''(r_m^*)}$: using Eq. (\ref{Gaus}), we obtain 
\begin{equation}
\rho ''(r_m^*)=\frac{\rho (r_m^*)}{(r_m^*)^2}\left[3-\frac{4.72}{\sigma ^2}\left(\frac{r_m^*}{r_m}\right)^2\right]
,\end{equation}
which, again for low $\sigma $ approximation, where $\rho (r_m^*)\approx \rho (r_m)$, we obtain
$\Delta r_m=-0.64\,\sigma ^2\,r_m$.

If we take the United Kingdom Infrared Deep Sky Survey (UKIDSS) data from \citet[Table 1]{Cab08} for in-plane lines of sight within $10^\circ <\ell <30^\circ $, $r_m$ derived from the maxima of star counts versus magnitude, an $\chi ^2$ analysis using Eq. (\ref{d_l}) including the errors
provided in the table, and $R_\odot =8$ kpc (as in \citealt{Cab08}), we obtain $\alpha =40.0^\circ \pm 1.7^\circ $, which is slightly lower than but compatible with the $\alpha =42.4^\circ \pm 2.1^\circ $ obtained by \citet{Cab08}, possibly because the fitting method is different.
However, when we applied the fit with $r_m^*$ (maxima of density versus distance) using Eq. (\ref{correction}) and the $\sigma $ values
given in the same table, we obtained $\alpha =27.4^\circ \pm 1.6^\circ $, which is similar to the result obtained by \citet{Weg15}
with UKIDSS and VISTA-VVV (Visible and Infrared Survey Telescope for Astronomy - Vista Variables in the V\'\i a L\'actea) data of 28--33$^\circ $ (a slightly larger value of $\alpha $ is possible due to a calibration of red clumps
with a slightly lower K-band absolute magnitude of -1.72 instead of the -1.62 of \citealt{Cab08}). 
For the $10^\circ <\ell <24^\circ $ range (avoiding the possible area of convergence with the 3 kpc arm), we find
$\alpha =24.7^\circ \pm 1.7^\circ $.
Similar considerations apply to the values obtained by \citet{Ham00}, \citet{Ben05}, \citet{Cab07}, and \citet{Dur16}. Therefore, a consensus value of around 25--35$^\circ $, similar to the inclination
of the triaxial bulge, is obtained.

\begin{figure}
\vspace{1cm}
{
\par\centering \resizebox*{8.5cm}{7cm}{\includegraphics{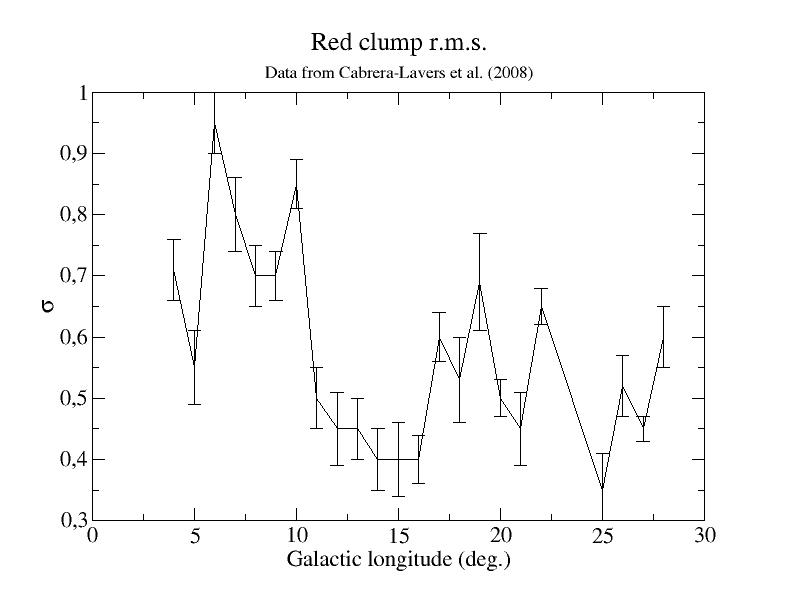}}
\par\centering
}
\caption{Dispersion of magnitudes in the peak of red clump giants derived by \citet{Cab08}.}
\label{Fig:sigma_Cab08}
\end{figure}

Another clue of what we observe in the red-clump distribution in the in-plane regions is given by the dispersion of magnitudes or
distances. \citet[Fig. 6]{Weg15} shows a distortion with respect to a straight long bar with $\alpha \approx 30^\circ $ at larger
longitudes ($\ell \gtrsim 20^\circ $), so the average distance of these stars is $5.5-6.0$ kpc instead of the $4.8-5.3$ kpc expected for $\alpha =28-33^\circ,   $ which was obtained by \citet{Weg15} as an average.
In a similar red-clump analysis by \citet{Cab08}, we see (in Fig. \ref{Fig:sigma_Cab08}) that the dispersion of magnitudes is $\lesssim 0.5$ mag for $10^\circ <\ell <17^\circ $, and for larger longitudes the peak gets much wider. The most direct interpretation
of these observations of the red clump is that the in-plane region with $20^\circ \lesssim \ell \lesssim 30^\circ $ contains a more complex density distribution: not only crossing a thin bar, but possibly a superposition of the long bar and another structure that spreads stars along much larger distances within the line of sight and pushes the average distance to higher values. 
This is possible, and indeed this second component is different from the thick bulge (limited to significant detection at $|\ell| \lesssim 10^\circ $); the long bar is a tangential cross of the 3 kpc arm, similar to the one observed at $\ell =-22^\circ$.

\begin{figure}
\vspace{1cm}
{
\par\centering \resizebox*{8.5cm}{8.5cm}{\includegraphics{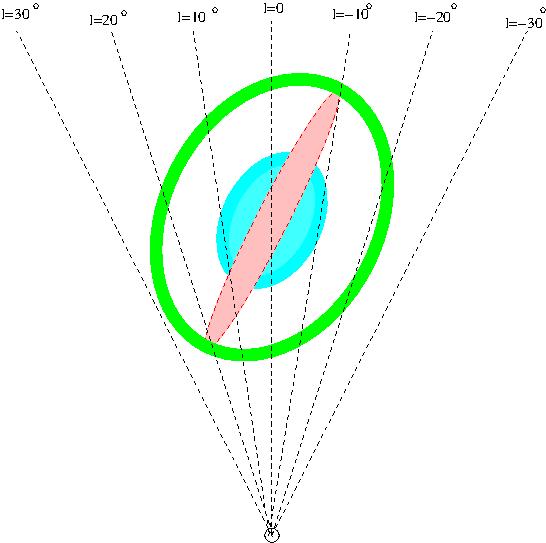}}
\par\centering
}
\caption{Graphical representation of the model explaining the in-plane observations discussed in this paper. Cyan shows the triaxial bulge; red shows the long bar; green stands for the 3 kpc arm. We set $\alpha =27^\circ $ for the three structures.}
\label{Fig:map}
\end{figure}

\begin{figure*}
\vspace{1cm}
{
\par\centering \resizebox*{17cm}{16cm}{\includegraphics{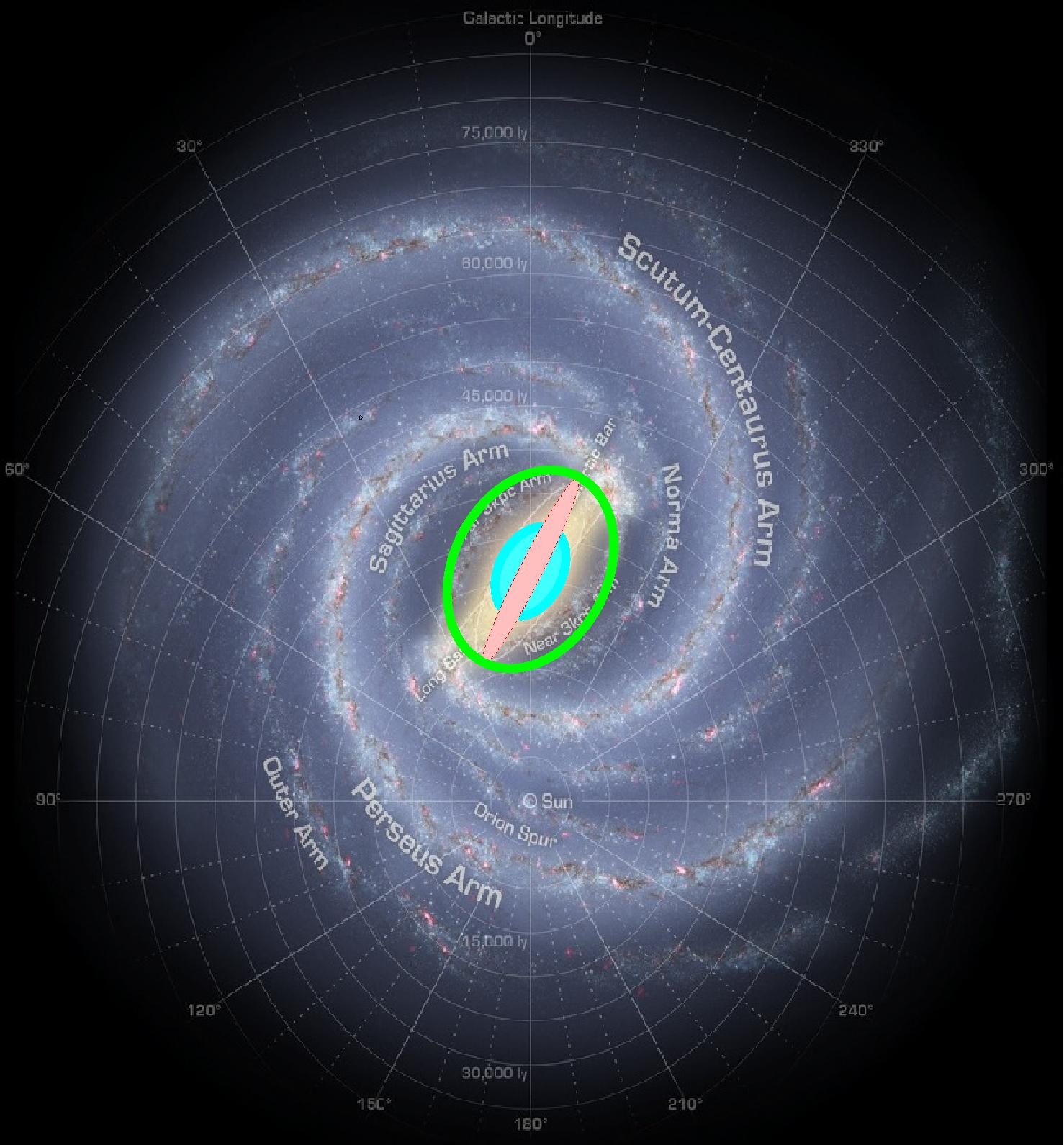}}
\par\centering
}
\caption{Superposition of model of Fig. \ref{Fig:map} with the
artistic representation of the annotated `Road Map to the Milky Way' by NASA/JPL-Caltech/R. Hurt (SSC/Caltech) 
based on NASA's \textit{Spitzer} Space Telescope achievements. Note that the angle of the long bar is amended.}
\label{Fig:map2}
\end{figure*}

\section{Discussion and conclusions}
\label{.discuss}

Figure \ref{Fig:map} gives a map of the structure in the inner in-plane Galaxy that
summarises our achievements from the previous sections: a model with a triaxial bulge plus a long bar and an ellipsoidal
ring (3 kpc arm), all of them with angles between the major axis and the line's Sun--Galactic centre of $\alpha =27^\circ $.
Figure \ref{Fig:map2} represents the amended representation of the whole Galaxy corresponding to the
artistic representation of the annotated `Road Map to the Milky Way' by NASA/JPL-Caltech/R. Hurt (SSC/Caltech) 
\footnote{https://www.spitzer.caltech.edu/image/ssc2008-10b-a-roadmap-to-the-milky-way-annotated} 
based on NASA's \textit{Spitzer} (Space Infrared Telescope Facility) Space Telescope achievements.

From our analysis, we cannot tell anything about the formation histories of the long bar or the bulge, including whether they are parts of the same structure formed at a given epoch or two different structures formed at two different epochs. Nonetheless, as mentioned in Sect. \ref{.intro}, 
the evidence presented in the literature supports the hypothesis that the bulge and the long bar 
are distinct structures;
the long bar has stars with different metallicities of the bulge, be they disc-like or higher \citep{Gon08,Weg19}, and younger ones than those of the bulge at a larger distance from the plane \citep{Ng96,Col02}. 
A recent analysis by \citet{Nie26} reveals a systematic age gradient, with a latitude
across the Galactic bulge within
$2^\circ <|b|<8^\circ $ shifting from a younger population with an average of $4.7^{+1.0}_{-0.8}$ Gyr prevalent near the plane to a predominantly older population with an average of $10.5^{+0.9}_{-0.8}$ Gyr at
higher latitudes. A young stellar population at low latitudes is predominantly composed of pseudo-bulges formed via disc/bar processes (incorporating contributions
from recent star-forming activity in the Galactic centre; \citealt{Nat17,Han25}), 
whereas the older stellar population
is associated with spheroidal bulges generated through early-stage collapse \citep{Obr13} or accretion of
debris from merged dwarf galaxies \citep{Obr13}.
Understanding the long bar as a wing-like extension of the bulge (a thicker structure) is possible too, 
but with a different period of formation. Calling these younger stars a 
`long bar' or a `peculiar extension of the bulge' is just a matter of semantics.

With this model, we can explain all of the previous observations. Said explanations are listed below.

\begin{itemize}
\item Star counts of in-plane bright sources with WISE-4.6$\mu $m magnitudes lower than 8 in $-10^\circ \lesssim \ell \lesssim 24^\circ $, either with extinction correction (Fig. \ref{Fig:WISEcountsce}) or without it (Fig. \ref{Fig:WISEcounts}), present a clear asymmetry between positive and negative longitudes. This is explained as
the projection of the long bar with the closer part in the first quadrant \citep{Lop07}. An angle as in Fig. \ref{Fig:map}, $\alpha \approx 27^\circ $, is compatible with the estimated angle taking into account the possible tips of the bar, 
$\alpha =36^{+13}_{-9}\ {\rm deg,}$ from Eq. (\ref{WISEalfa}). In particular, using Eq. (\ref{barangletips}),
we can obtain a value of $\alpha =27^\circ $ if
we place the tips at $\ell \approx -9^\circ$, and $\ell =24^\circ $. With these angles, the semi-axis radius of the long-bar is $L_0=4.3$ kpc. 
One or two degrees of broadening in the tips may be observed due to the thickness of the structure.

\item For the long bar within $10^\circ <\ell <24^\circ $, the value of $\alpha $ of the weighted average of the results of the different observations with distance information analysed in this paper is (27.4$\pm $1.5) deg; the average of
APOGEE M giants with \citet{Sto24} distances is (25.5$\pm $2.3) deg (Sect. \ref{.APOGEEangle}); the average of O-rich ones with ages of $<5$ Gyr (Sect. \ref{.Mirasalfa}) is (28.8$\pm 2.0$) deg.

\item The distance distribution of near-plane APOGEE M giants (Fig. \ref{Fig:counts}) or red clumps (K giants) in the literature
(see Sect. \ref{.rc}) would observe the long bar too, with a similar angle to Fig. \ref{Fig:map}. 
Previous determinations of larger values of the angle $\alpha $ are due to miscalibration of standard candles or the determination of the peak of red clumps in the star counts versus magnitude instead of the density versus distance. The higher spread of
distances at $20^\circ \lesssim \ell \lesssim 30^\circ $ (Fig. \ref{Fig:sigma_Cab08}) would be at the intersection of the long
bar with tangential cut of the stellar ring, with an increasing average distance for $24^\circ \lesssim \ell \lesssim 28^\circ $.

\item In-plane Miras variable stars with ages of $\lesssim $5 Gyr (Figs. \ref{Fig:MIRASdist} and \ref{Fig:MIRAScart}) show 
a remarkable similarity with the proposed scenario. We see the presence
of the triaxial bulge within $|\ell |\lesssim 10^\circ $ and its extension of the long bar until $\ell =+24^\circ $.
We also very clearly see the imprint of the tangential cut of the stellar ring at $\ell \approx -22^\circ $, which is separated from the bar at negative longitudes and the equivalent point at positive longitudes at $\ell \approx 27^\circ $; it is at a shorter distance from the 
tip of the bar, in agreement with previous works (\citet{Mik82,Rue91}). 
Therefore, with this scenario, the tip of the bar in positive longitude would
not be at $\ell =-27^\circ $ as previously posited \citep{Lop99,Lop07}, but at $\ell \approx 24^\circ $.
The proximity of the tip of the bar to the tangential cut of the ring gives place to an almost continuity of
stars without any  (or with a very small) gap between them. Moreover, the bottom of Fig. \ref{Fig:MIRASdist}
shows an almost continuous structure at a heliocentric distance of $4.0-6.0$ kpc within $-15^\circ \lesssim \ell \lesssim 30^\circ, $
corresponding to the closest part of the near 3 kpc arm, whereas at  $-25^\circ \lesssim \ell \lesssim -15^\circ $ we see
a much higher spread of heliocentric distances with 5-12 kpc. There is a possible continuation towards the far 3 kpc arm that is
tentatively detected beyond the bulge and long bar at certain distances (light green in the bottom part of Fig. \ref{Fig:MIRASdist}
at $-22^\circ \lesssim \ell \lesssim +27^\circ $; distances are 9--12 kpc).

\end{itemize}

In a model of four spiral arms, the peaks in the star counts of the old population as a function
of Galactic longitude are explained as \citep[Table 8]{Val22} $\ell =-22^\circ $ being the start of Perseus Arm; $\ell =-12^\circ $ 
as the start of Sagittarius Arm; $\ell =19^\circ $ as the start of Norma Arm; and $\ell =31^\circ $ as the tangential point of the Scutum Arm.
However, this \citet{Val17,Val22} model cannot explain some observed features: the almost constant distance of 5.0--5.5 kpc for
O-rich Miras between $\ell =19^\circ $ and $\ell =27^\circ $ (Fig. \ref{Fig:MIRAScounts}) or the continuous M giants'
star counts  between $\ell =19^\circ $ and $\ell =24^\circ $ (Fig. \ref{Fig:WISEcountsce}). Moreover, spiral arms usually have
very young stellar populations, whereas our observations show a significant presence of an old population, either in terms of red clumps (K giants), M giants, or Miras with ages of $\lesssim $5 Gyr. Therefore, rather than spiral arms, the presence of the
3 kpc arm (which, in spite of the name, is not a spiral arm, but something of a different nature and part of older stellar populations) 
better fits the observations.

We solved several controversies concerning the in-plane central area of the Milky Way: an alignment with $\alpha \approx 27^\circ$ between the bulge and the long bar and the 3 kpc arm's major axis explains everything without the need to introduce different angles for each structure. We also tentatively detect stars from the far 3 kpc arm, and we consistently clarified the nature of the excess of stars at $\ell =-22^\circ $ and $\ell =+27^\circ $ in terms of the tangential cuts of lines of sight with the ellipsoid of the 3 kpc arm. The long bar has a semi-axis radius of around 4 kpc, which is also the size of the major axis of the 3 kpc arm (minor axis $\approx 3$ kpc).
\\

\begin{acknowledgements}
Thanks are given to the anonymous referee for helpful comments and suggestions.
Thanks are given to the A\&A language editor Natasha Saint Geni\`es.
MLC and FGL acknowledge support from the Spanish Ministerio de
Ciencia, Innovación y Universidades (MICINN) under grant number PID2021-129031NB-I00.

This publication makes use of data products from the Wide-field Infrared Survey Explorer, which is a joint project of the University of California, Los Angeles, and the Jet Propulsion Laboratory/California Institute of Technology, and NEOWISE, which is a project of the Jet Propulsion Laboratory/California Institute of Technology. WISE and NEOWISE are funded by the National Aeronautics and Space Administration. The AllWISE program is funded by the NASA Science Mission Directorate Astrophysics Division. AllWISE data products are generated using the imaging data collected and processed as part of the original WISE and NEOWISE programs.

Funding for the Sloan Digital Sky 
Survey IV has been provided by the 
Alfred P. Sloan Foundation, the U.S. 
Department of Energy Office of 
Science, and the Participating 
Institutions.  SDSS-IV acknowledges support and 
resources from the Center for High 
Performance Computing  at the 
University of Utah. The SDSS 
website is www.sdss4.org.
SDSS-IV is managed by the 
Astrophysical Research Consortium 
for the Participating Institutions 
of the SDSS Collaboration including 
the Brazilian Participation Group, 
the Carnegie Institution for Science, 
Carnegie Mellon University, Center for 
Astrophysics | Harvard \& 
Smithsonian, the Chilean Participation 
Group, the French Participation Group, 
Instituto de Astrof\'isica de 
Canarias, The Johns Hopkins 
University, Kavli Institute for the 
Physics and Mathematics of the 
Universe (IPMU) / University of 
Tokyo, the Korean Participation Group, 
Lawrence Berkeley National Laboratory, 
Leibniz Institut f\"ur Astrophysik 
Potsdam (AIP),  Max-Planck-Institut 
f\"ur Astronomie (MPIA Heidelberg), 
Max-Planck-Institut f\"ur 
Astrophysik (MPA Garching), 
Max-Planck-Institut f\"ur 
Extraterrestrische Physik (MPE), 
National Astronomical Observatories of 
China, New Mexico State University, 
New York University, University of 
Notre Dame, Observat\'ario 
Nacional / MCTI, The Ohio State 
University, Pennsylvania State 
University, Shanghai 
Astronomical Observatory, United 
Kingdom Participation Group, 
Universidad Nacional Aut\'onoma 
de M\'exico, University of Arizona, 
University of Colorado Boulder, 
University of Oxford, University of 
Portsmouth, University of Utah, 
University of Virginia, University 
of Washington, University of 
Wisconsin, Vanderbilt University, 
and Yale University.
Collaboration Overview
Affiliate Institutions
Key People in SDSS
Collaboration Council
Committee on Inclusiveness
Architects
SDSS-IV Survey Science Teams and Working Groups
Code of Conduct
SDSS-IV Publication Policy
How to Cite SDSS
External Collaborator Policy
For SDSS-IV Collaboration Members

We acknowledge the use of OGLE survey, a Polish astronomical project based at the 
University of Warsaw that runs a long-term variability sky survey (1992--present). Most of 
the observations have been made at the Las Campanas Observatory in Chile. Cooperating institutions include 
Princeton University and the Carnegie Institution.
\end{acknowledgements}

\appendix
\section{Fit of ellipse parameters}
\label{.ellipse}

\begin{figure}
\vspace{1cm}
{
\par\centering \resizebox*{6cm}{9.8cm}{\includegraphics{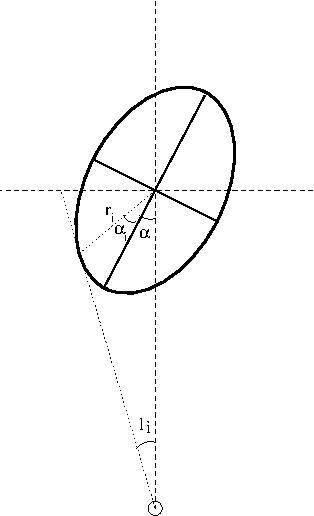}}
\par\centering
}
\caption{Graphical representation of an ellipse with a tangential point of a line of sight from the Sun.}
\label{Fig:ellipse}
\end{figure}

An ellipse has three free parameters: the major axis $a$, the minor axis $b$ and the angle of the major axis $\alpha $.
It follows
\begin{equation}
\label{elipse}
y_{e,i}=\pm \frac{b}{a}\sqrt{a^2-x_{e,i}^2}
\end{equation}\[
r_{i}=\sqrt{x_{e,i}^2+y_{e,i}^2}
,\]\[
\tan (\alpha _i)=\frac{y_{e,i}}{x_{e,i}}
,\]
where $(x_{e,i},y_{e,i})$ are coordinates of a point $i$ of the ellipse in the reference system of the ellipse ($x_e$: 
major axis; $y_e$: minor axis).
For each tangential point $i=1,2$ corresponding to lines of sight with Galactic longitude $\ell _i$ (see Fig. \ref{Fig:ellipse}),
we have the direction tangential to the ellipse with angle 
$\alpha _{t,i}=\frac{dy_{e}}{dx_{e}}(x_{e,i})$ (in the system of the ellipse), 
parallel to the line of sight with Galactic longitude $\ell _i$ (in Galactocentric coordinates, with X-axis crossing
the Sun), and thus
\begin{equation}
\tan (\alpha _i)=\left(\frac{b}{a}\right)^2\frac{1}{\tan (\alpha +\ell _i)}
.\end{equation}
Also, this tangential point follows the sine rule
\begin{equation}
r_{i}=\frac{R_\odot \sin (\ell _i)}{\sin (\alpha +\alpha _i+\ell _i)}
.\end{equation}
Another information is the distance $d_3$ of the ellipse at $\ell _3=0$, $\alpha _3=-\alpha $
thus we have a third point $i=3$ with
\begin{equation}
r_{3}=R_\odot -d_3
.\end{equation}

Therefore, we have five equations (for $r_{e,i}$, $i=1,2,3$; $\alpha _i$, $i=1,2$) that allow five independent parameters to be determined -- three parameters of the ellipse ($\alpha $, $a$, $b$) and the two angles of the tangential points $\alpha _i$, $i=1,2$ -- from the five observed numbers $\ell _i$, $i=1,2,3$; $R_\odot $, $d_3$. The angle $\alpha _3$ is not independent, but equal to $\alpha $; the values of the radii $r_i$ are not independent, but related to $\alpha _i$ through Eqs. (\ref{elipse}).

\end{document}